\documentclass[final,twocolumn,12pt]{elsarticle}

\usepackage{epsfig}

\usepackage{amssymb}
\usepackage[latin1]{inputenc}
\usepackage[T1]{fontenc}

\journal{Journal of Non-Crystalline Solids}

\begin{document}

\begin{frontmatter}



\title{Structure and dynamics of liquid AsSe$_4$ from \textit{ab initio} molecular dynamics simulation}


\author{M. Bauchy}

\address{Laboratoire de Physique Théorique 
de la Matière Condensée, Université Pierre et Marie Curie, Boite 121,
4, place Jussieu 75252 Paris Cedex 05, France}

\begin{abstract}

Structural and dynamical properties of AsSe$_4$ liquids have been studied by \textit{ab initio} molecular dynamics simulations as a function of temperature. Calculated neutron structure factors are in good agreement with experimental data. Results show the existence of a significant amount of As-As homopolar bonds and of Se$^{\rm I}$ and As$^{\rm IV}$ units which are not part of the picture of the cross-linking As(Se$_{1/2}$)$_3$ pyramids usually proposed for the glassy state.

\end{abstract}


\end{frontmatter}


\section{Introduction}
\label{sec:intro}


As a typical amorphous semiconductor, As$_x$Se$_{1-x}$ glasses and liquids have been extensively studied for more than 40 years since the pioneering work of Meyers and Felty \cite{felty_preparation_1965}. For $x$<2/5, their structure is usually described as a random network of Se$_q$ chains cross-linked by pyramidal As(Se$_{1/2}$)$_3$ units \cite{felty_preparation_1965}. This model is supported by the observed local maximum of $T_g(x)$ near $x$=2/5, when the cross-linking process saturates, pyramidal units being connected to each other by only one Se atom.


Unlike silicate systems that have been widely simulated via classical molecular dynamics, the electronic structure of chalcogenide systems has to be explicitly treated to account for charge transfers between atoms. The use of \textit{ab initio} molecular dynamics is therefore required, thus inducing the use of systems of only hundreds of atoms simulated over a few ps. Several studies have been reported, most of them focusing on the stoichiometric As$_2$Se$_3$ system \cite{shimojo_microscopic_1999, shimojo_temperature_2000} or on the glassy state \cite{li_first-principles_2000, drabold_simulations_2003}. Recently, several compositions in the liquid state were simulated by Zhu \cite{zhu_ab_2008}. It should also be stressed that many chalcogenides contain homopolar defects \cite{salmon_structure_2003} that can only be reproduced from a careful electronic modelling using density functional theory \cite{micoulaut_improved_2009, massobrio_computer_2010} whereas such features are usually not reproduced from classical molecular dynamics \cite{mauro_multiscale_2007}.


In this paper, we focus on the low As amount $x$=0.20 composition and use a larger system than previous simulations \cite{zhu_ab_2008} to get enough statistics to study the environment of As atoms. After giving the simulation methodology in Sec. \ref{sec:methods}, results are presented in Sec. \ref{sec:results} and discussed in Sec. \ref{sec:discussion}. Finally, conclusions are given in Sec. \ref{sec:conclusion}.

\section{Methodology}
\label{sec:methods}

First principles simulations \cite{car_unified_1985} were performed to simulate AsSe$_4$ liquids of 200 atoms at different temperatures (600, 800, 1200, 1600 and 2000~K). A cubic box with periodic boundary conditions was used, its length being fixed in order to recover the ambient density of the glass \cite{georgiev_rigidity_2000}. This methodology induces the existence of pressure at high temperature (0.32~GPa at 2000K), but the latter remains small as compared to pressure fluctuations in the box (0.5 GPa at 2000K). The electronic structure was described within density functional theory. Valence electrons were treated explicitly, in conjunction with norm-conserving pseudopotentials to account for core-valence interactions \cite{troullier_efficient_1991}. The wave functions were expanded at the $\Gamma$ point of the supercell and the energy cutoff was set at 20 Ry. Other parameters of the simulation (fictitious mass, time step, exchange-correlation scheme, GGA) are identical to the ones used in previous simulations on GeSe$_2$ and Ge-Se liquids and glasses \cite{micoulaut_improved_2009, massobrio_impact_2010, bauchy_angular_2011}. All temperature points were accumulated over 20~ps.

\section{Results}
\label{sec:results}

\subsection{Structure factor}

To compare the structure of the simulated liquids with experimental and previous simulations results, we computed neutron structure factors. The partial structure factors have been first calculated from the pair distribution functions (PDFs) $g_{ij}(r)$ :

\begin{equation}
S_{ij}(Q) = 1 + \varrho_0 \int_{0}^R 4\pi r^2 (g_{ij}(r)-1) \frac{\sin (Qr)}{Qr} F_{L}(r)\, \mathrm dr
\end{equation} where $Q$ is the scattering vector, $\varrho_0$ is the average atom number density and $R$ is the maximum value of the integration in real space (here $R = 8$~\AA). The $F_{L}(r) = \sin (\pi r / R) / (\pi r / R)$ term is a Lortch-type window function used to reduce the effect of the finite cutoff of $r$ in the integration \cite{wright_neutron_1988}. As discussed in \cite{du_compositional_2006}, the use of this function reduces the ripples at low $Q$ but induces a broadening of the structure factor peaks. The total neutron structure factor can then be evaluated from the partial structure factors following :

\begin{equation}
S_N(Q) = \bigl( \sum_{i,j=1}^n c_ic_jb_ib_j \bigr) ^{-1} \sum_{i,j=1}^n c_ic_jb_ib_j S_{ij}(Q)
\end{equation} where $c_i$ is the fraction of $i$ atoms (As, Se) and $b_i$ is the neutron scattering length of the species (given by 6.58 and 7.97~fm for arsenic and selenium atoms respectively \cite{zhu_ab_2008}).

\par

\begin{figure}[t]
\begin{center}
\includegraphics[width=\linewidth]{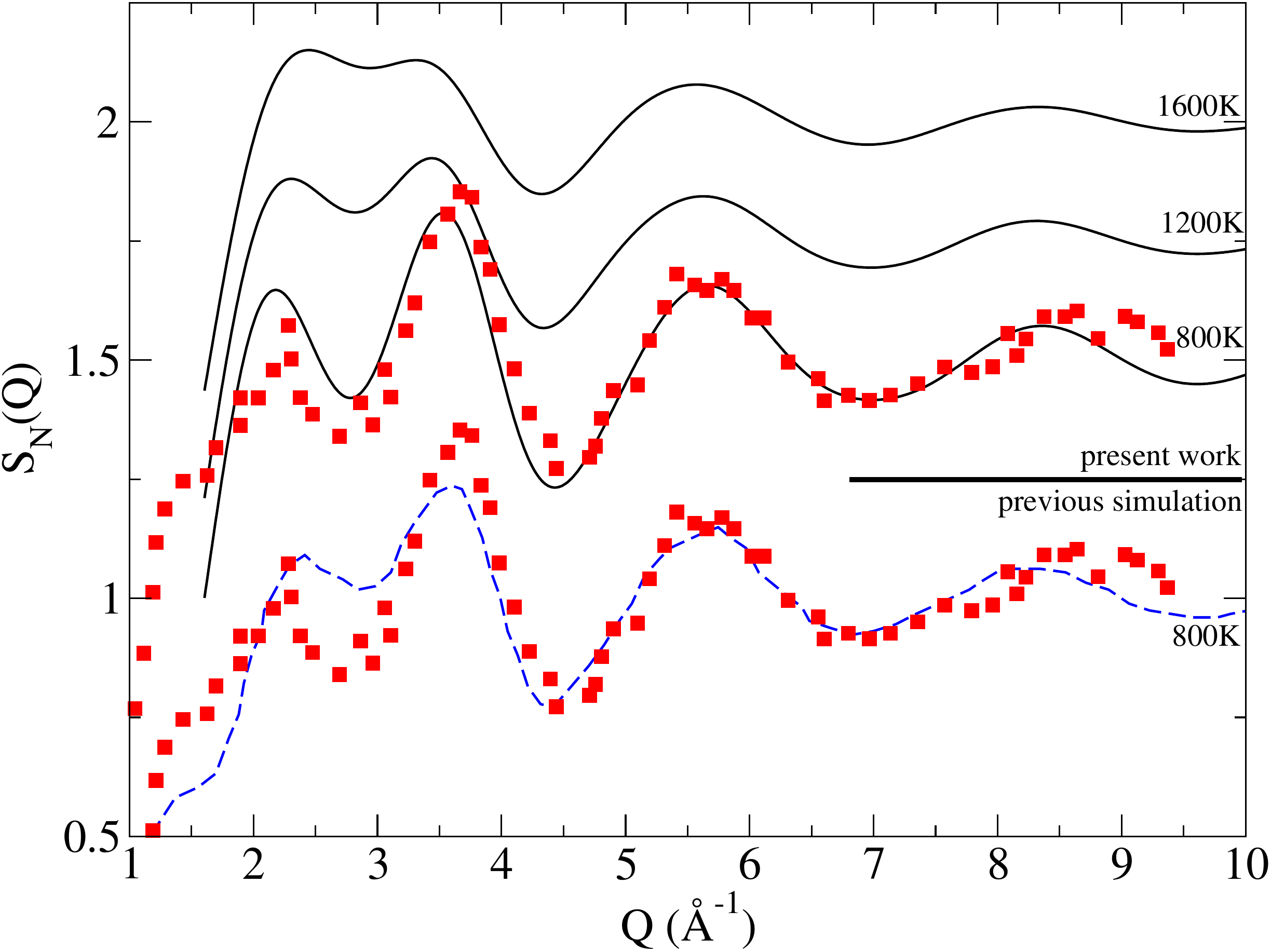}
\end{center}
\caption{\label{fig:S_as40} Calculated neutron structure factors S$_N$(Q) of liquid AsSe$_4$ (top, black solid lines) at different selected temperatures, compared with a previous simulation from Zhu (bottom, blue broken line, \cite{zhu_ab_2008}). Calculated structure factors at 800~K are compared with neutron diffraction data from Usuki (red squares \cite{usuki_neutron_1998}). (color online)}
\end{figure}

Neutron structure factors of liquid AsSe$_4$ are represented on Fig. \ref{fig:S_as40} for different selected temperatures and compared with a previous simulation from Zhu \cite{zhu_ab_2008} and neutron diffraction data from Usuki \cite{usuki_neutron_1998}. We observe that the agreement with experimental data for AsSe$_4$ is improved in the present simulation as compared to the one of Zhu \cite{zhu_ab_2008}. This may be due to the fact that larger systems are required to get reasonable statistical averages at low wave vector since, for this composition, the system of Zhu \cite{zhu_ab_2008} contains only 20 As atoms as compared to 40 in the present simulation.

\par

The good agreement with experiments of the simulated AsSe$_4$ liquid at 800~K allows us to study temperature effects. As observed in experiments \cite{uemura_structure_1978, hosokawa_x-ray-diffraction_1992, kajihara_x-ray_2007, crozier_structural_1977}, the intensity of the first peak of S$_N$(Q) at 2.3~\AA$^{-1}$ increases and its width increases slightly with the temperature while the two first peaks progressively merge each other to form a broad one over 2-4~\AA$^{-1}$.

\subsection{Pair distribution functions}

\begin{figure}[t]
\begin{center}
\includegraphics[width=\linewidth]{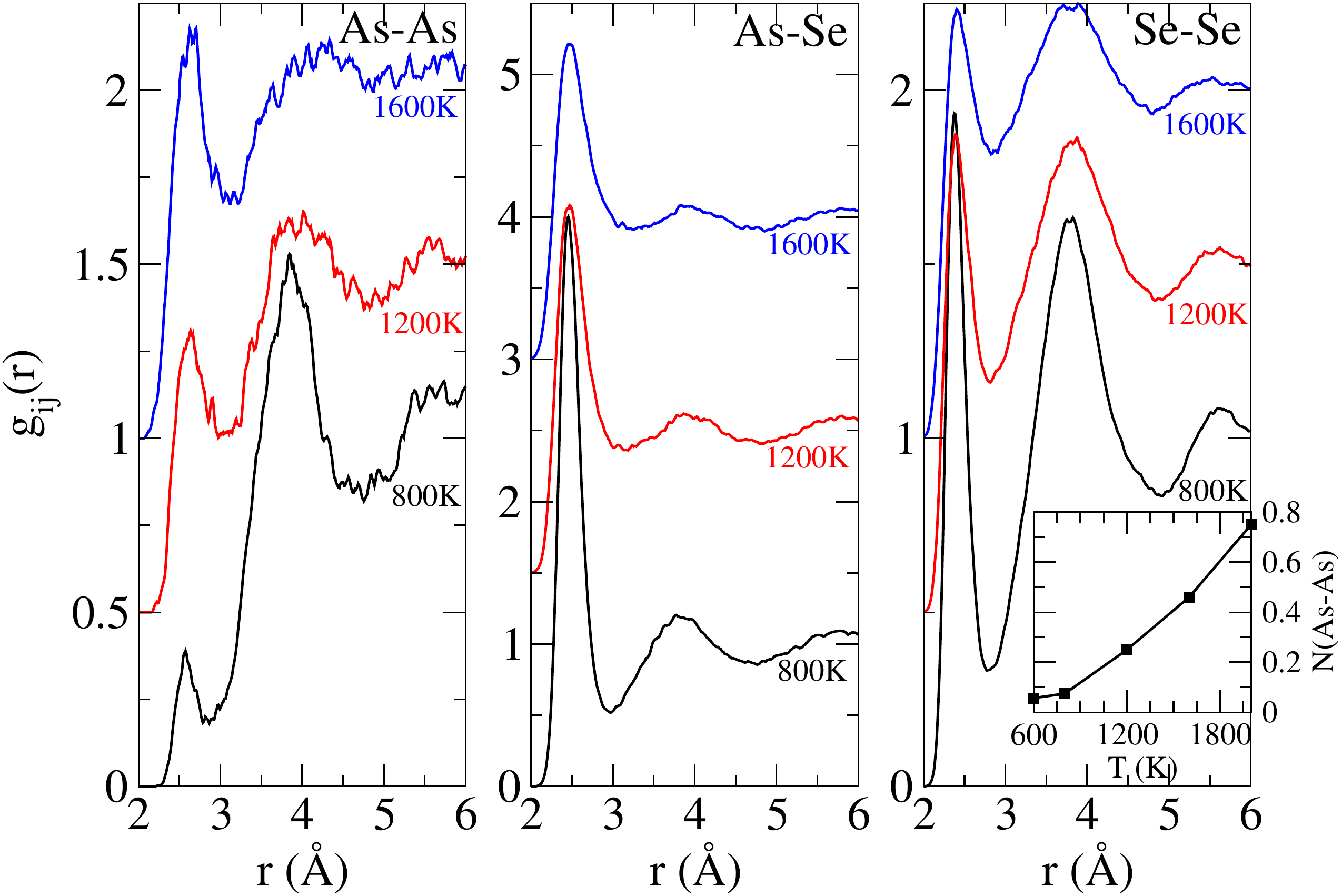}
\end{center}
\caption{\label{fig:gij_as20} Partial pair distribution functions (PDFs) g$_{ij}$(r) As-As, As-Se and Se-Se of AsSe$_4$ liquids at different selected temperatures. The inset shows the As-As partial coordination number N$_{\rm AsAs}$ according to the temperature. (color online)}
\end{figure}

The local structural order can be observed in more details by considering the partial PDFs g$_{ij}$(r). As-As, As-Se and Se-Se partial of AsSe$_4$ liquids are represented on Fig. \ref{fig:gij_as20} at different temperatures. As temperature increases, the usual trends are observed, manifested by a decrease of the intensity and a broadening of the peaks. We find that large changes in structure take place between 800~K and 1200~K which manifest by a rapid decrease of the main peaks (3.7, 2.5 and 3.8~\AA~for As-As, As-Se and Se-Se respectively). For As-Se and Se-Se, these typical length scales correspond to the distances defining the As(Se$_{1/2}$)$_3$ pyramid. The variation shows that the latter must be substantially modified in this temperature rang as discussed below.

\subsection{Coordination numbers}

As presented earlier, the average coordination numbers (CNs) of As and Se atoms can be calculated via the integration of the first peak of the partial PDFs. After respective integrations to the first minimum of each PDF, we obtain N$_{\rm AsAs}$~=~0.075, N$_{\rm AsSe}$~=~2.95, N$_{\rm SeAs}$~=~0.74 (note that, as implied by the stoichiometry, N$_{\rm AsSe}$~=~4N$_{\rm SeAs}$ ) and N$_{\rm Se-Se}$~=~0.94 for $T$~=~800~K so that N$_{\rm As}$~=~3.03 and N$_{\rm Se}$~=~1.68. At 1600~K, these values change to N$_{\rm AsAs}$~=~0.46, N$_{\rm AsSe}$~=~2.77, N$_{\rm SeAs}$~=~0.69 and N$_{\rm Se-Se}$~=~0.72.

\par

\begin{figure}[t]
\begin{center}
\includegraphics[width=\linewidth]{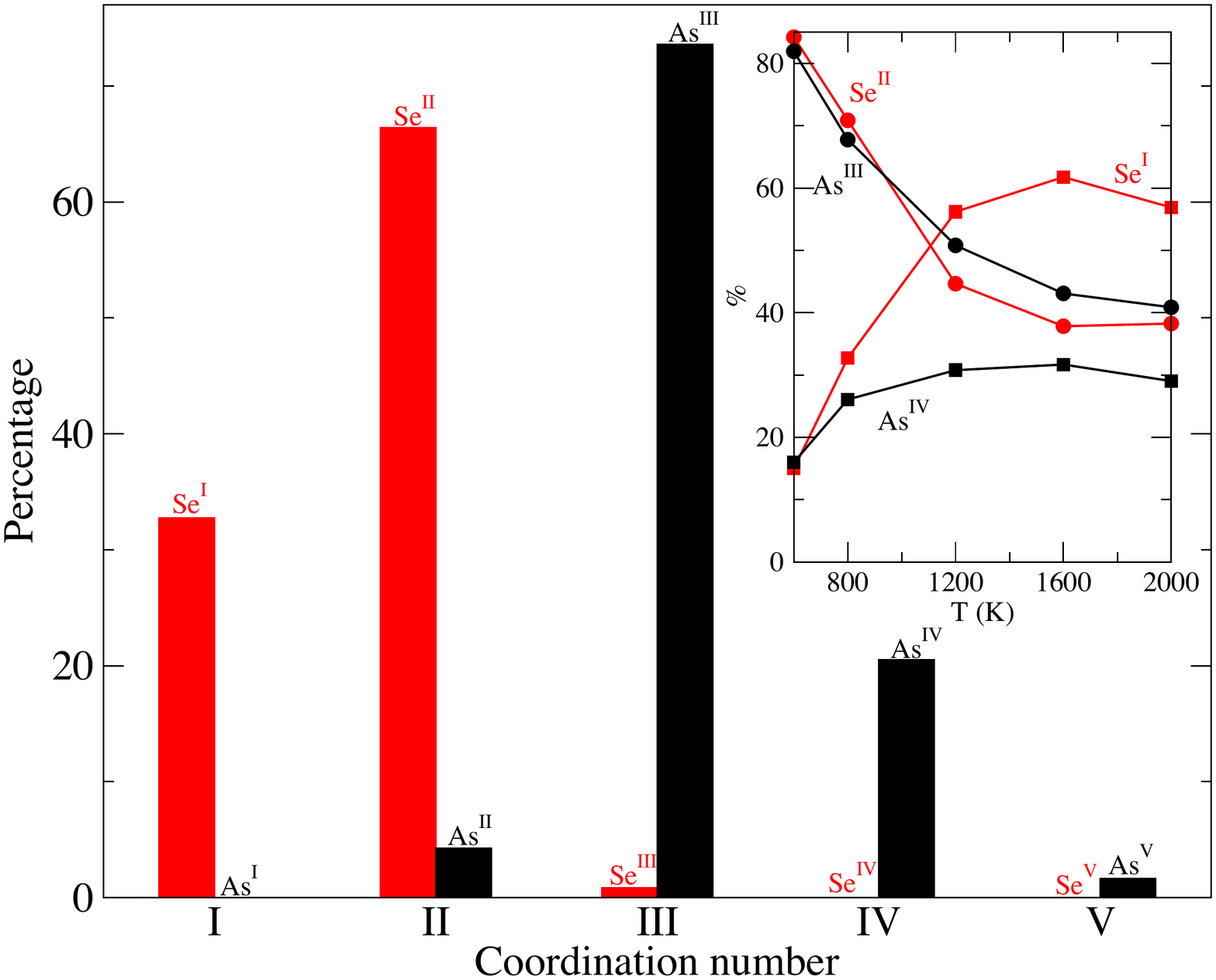}
\end{center}
\caption{\label{fig:n-as40} Coordination numbers (CNs) populations of As and Se atoms in AsSe$_4$ at 800~K. The inset shows the behavior of these coordination numbers populations in AsSe$_4$ with respect to the temperature and (right) the evolution of the percentage of As$^{\rm IV}$ and Se$^{\rm I}$ populations in As$_x$Se$_{1-x}$ liquids at 800~K according to $x$. (color online)}
\end{figure}

However, to get the full distribution of the different coordinated species, it is useful to enumerate the number of neighbors that are present in the first coordination shell of each atom at each step. The populations of the different CNs in the liquid AsSe$_4$ at 800~K are shown on Fig. \ref{fig:n-as40}. As expected, the dominant species are the base units of the cross-linking pyramid model, Se$^{\rm II}$ and As$^{\rm III}$. However, a significant amount of defect Se$^{\rm I}$ and As$^{\rm IV}$ units are found (respectively 32.7\% and 20.5\%), as well as a small amount of Se$^{\rm III}$ (0.9\%) and As$^{\rm V}$ atoms (1.6\%). Other contributions (Se$^{\rm 0}$, Se$^{\rm IV}$, As$^{\rm I}$ and As$^{\rm VI}$) are insignificant (<1\%).

\par

The inset in Fig. \ref{fig:n-as40} shows the evolution of the main species in AsSe$_4$ with respect to the temperature. We notice that only minor changes in CNs are observed until the temperature becomes lower than 1200~K. As temperature continuously decreases, the fractions of Se$^{\rm I}$ and As$^{\rm IV}$ units starts to decrease while the ones of Se$^{\rm II}$ and As$^{\rm III}$ increase.

\subsection{Diffusion}

In order to analyze the effect of the temperature on the dynamics of AsSe$_4$, the mean square displacements (MSD) $\langle r^2(t) \rangle_{\alpha}$ ($\alpha$~=~Se, As) of Se and As atoms have been computed according to~:

\begin{equation}
\langle r^2(t) \rangle_{\alpha} =\frac{1}{N_{\alpha}} \sum_{i=1}^{N_{\alpha}}
\langle |{\bf r}_i(t)-{\bf r}_i(0)|^2\rangle\quad,
\end{equation} where $N_{\alpha}$ is the number of $\alpha$ atoms and ${\bf r}_i(t)$ the position of the atom $i$. The diffusion constants D$_{\alpha}$ have then been calculated from the long-time limit of the MSD~:

\begin{equation}
\label{eq:D}
D_{\alpha} = \lim_{t\to \infty} \langle r^2(t)\rangle_{\alpha}/6t
\end{equation}

\par

\begin{figure}[t]
\begin{center}
\includegraphics[width=\linewidth]{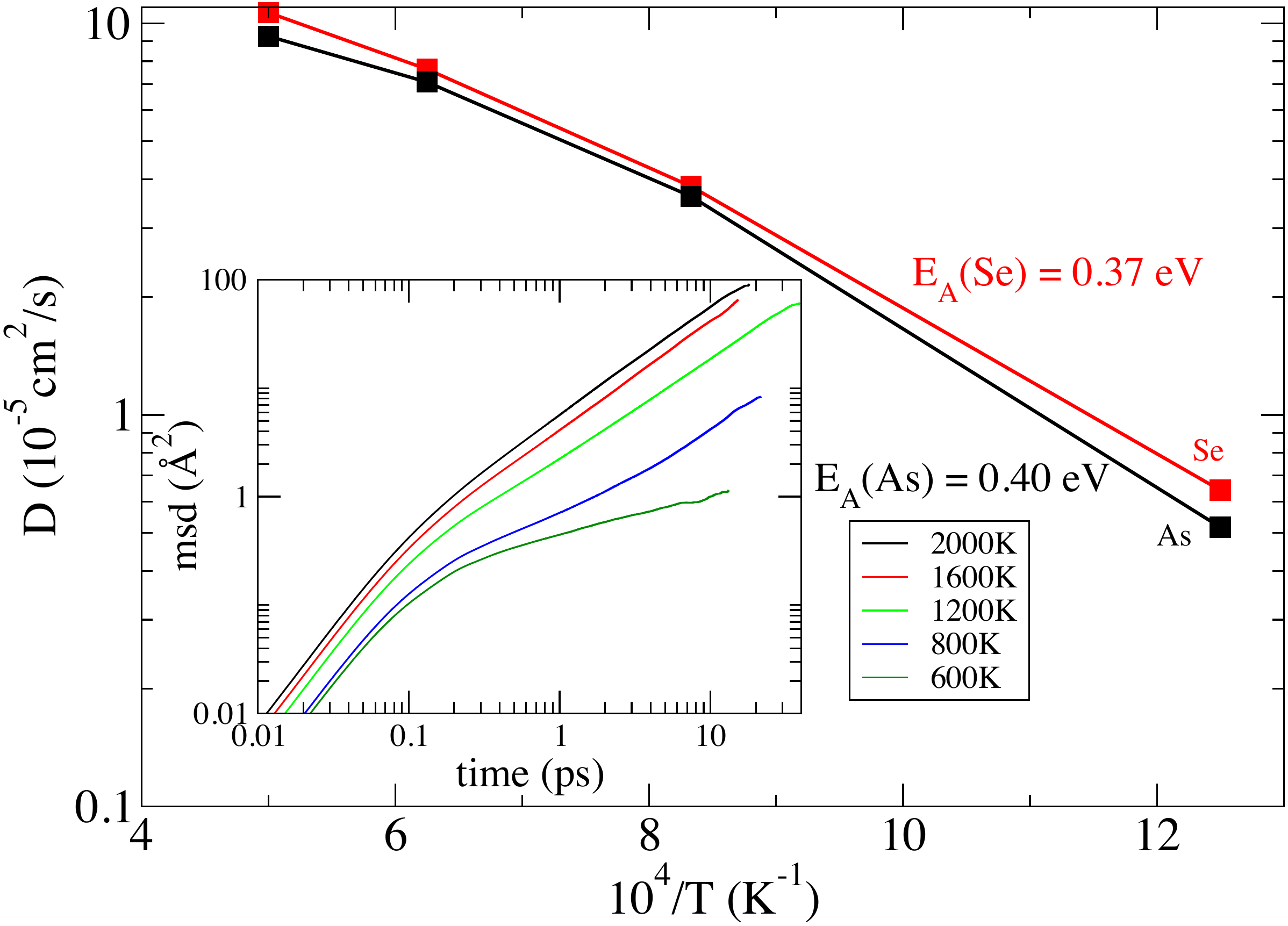}
\end{center}
\caption{\label{fig:D_as40} Self-diffusion constant of Se and As atoms in AsSe$_4$ liquids at different temperatures. Associated action energies E$_A$ are mentioned. The inset shows the mean square displacement of Se atoms at different temperatures. (color online)}
\end{figure}

The inset in Fig. \ref{fig:D_as40} shows the MSD of Se atoms at different selected temperatures. At high temperature (1600~K), the diffusive regime, which manifests by a slope of 1 in a log-log plot of the MSD with time, onsets at about 0.3~ps. This regime can be clearly observed at all temperatures except at 600~K since, due to the slowing down of the dynamics as the temperature decreases, the diffusive regime is not reached over the simulation time (20~ps). The associated diffusion constants, computed from Eq. \ref{eq:D}, are shown in Fig. \ref{fig:D_as40}. To the best of our knowledge, no experimental nor simulation data are available for As-Se liquids. A non-Arrhenius behavior is obtained at high temperature, which is also found in oxide glasses \cite{micoulaut_simulated_2006}. Although the number of data points is limited, the obtained diffusion constants have been fitted by an Arrhenius law~:
\begin{equation}
D_{\alpha} = \exp \bigl(- \frac{E_A}{kT} \bigr)
\end{equation} in order to extract the activation energies $E_A$(Se) and $E_A$(As) which are respectively equal to 0.37 and 0.40~eV. These values are close to those determined in Ge-Se liquids (0.43~eV for the diffusion of Se atoms in GeSe$_4$ but 1.1~eV for GeSe$_2$ \cite{micoulaut_improving_2009}), but certainly much smaller than those found in oxides (for example 1.12~eV for the diffusion of O atoms in GeO$_2$ \cite{micoulaut_simulated_2006}). The reason of this difference may arise from the fact that one has a low connected system where homopolar Se$_q$ chains dominate which should favor diffusion across the liquid. This seems in line with recent studies showing that flexible (i.e. weakly connected) networks display an increased diffusion or ionic conductivity \cite{novita_fast-ion_2007, micoulaut_fast-ion_2009}.


\section{Discussion}
\label{sec:discussion}

As it can be seen on Fig. \ref{fig:gij_as20}, the existence of a first peak in the partial As-As shows that homopolar bonds As-As do exist at all temperatures, even if they become more predominant at high $T$, so that the usual cross-linking pyramid model is not realistic in the liquid state. Note that these As-As bonds have been observed by EXAFS measurements \cite{tamura_exafs_1992} in As$_2$Se$_3$. The behavior of the As-As partial coordination number is shown in the inset of Fig. \ref{fig:gij_as20}. Its value at 800~K (0.075) is found to be lower than the one of Zhu (0.12 \cite{zhu_ab_2008}). The increase of the fraction of As-As bonds with the temperature has been observed experimentally by Hosokawa et al. \cite{tamura_exafs_1992}.

\par

Moreover, the significant amount of As$^{\rm IV}$ and Se$^{\rm I}$ species also contradicts the latter model and may be the signature of the existence of a Se=As(Se$_{1/2}$)$_3$ basic unit. the existence of the latter has been suggested by Georgiev et al. \cite{georgiev_rigidity_2000} using an approach based on topological constraints in As-Se glasses and observed experimentally in the corresponding sulfide As-S glass \cite{wagner_glass_1996, diemann_amorphous_1979}. However, we note that, for As-Se systems, no clear experimental evidence of the existence of this unit from direct structural analysis has been reported so far. We note that no As$^{\rm IV}$ were found in Shimojo's simulations \cite{shimojo_temperature_2000} in As$_2$Se$_3$. The average CN of As atoms found by Usuki by neutron diffraction in AsSe$_4$ (3.01 $\pm$ 0.02 \cite{usuki_neutron_1998}) is in very good agreement with our value (3.03 as compared to 2.67 in Shimojo's simulation) and suggests a minor fraction of As$^{\rm IV}$ atoms in the liquid phase.

\section{Conclusions}
\label{sec:conclusion}

We have carried out \textit{ab initio} molecular dynamics simulations in AsSe$_4$ liquids in order to study the effect of temperature on the structure and dynamics. Thanks to the use of a larger system, the calculated neutron structure factors show a better overall agreement with experimental data as compared to previous simulations. Results show that the cross-linking pyramid model is not suitable any more in the liquid state and that exotic Se$^{\rm I}$ and As$^{\rm IV}$ units exist as well as As-As bonds. In future works, we plan to check if these anomalies can still be found in the glassy state and in other compositions.

\section*{Acknowledgments}
\label{sec:acknowledgments}

Warm thanks are due to M. Micoulaut for suggesting this study and providing advice at its various stages and to A. Kachmar and P. Boolchand for stimulating discussions.





\end{document}